\documentstyle[prl,multicol,aps,epsf]{revtex}
\begin{document}
\title{Spin-One Bosons in 
Low Dimensional Mott Insulating States}
\author{Fei Zhou}
\address{ITP,Utrecht University, Minnaert building, Leuvenlaan 4,
3584 CE Utrecht, The
Netherlands}
\maketitle

\begin{abstract} 
We analyze the strong coupling limit of spin-one bosons in
low dimensional lattices. In one-dimensional lattices, for an odd number
of bosons per site ($N_0=2k+1$), the ground state is a dimerized valence 
bond
crystal with a two-fold degeneracy; the low lying elementary spin
excitations carry spin-one. For an even number of bosons per site 
($N_0=2k$), the
ground state is a nondegenerate spin singlet Mott state. We also argue
that in square lattices 
in a quantum spin disordered limit, ground states 
should be dimerized valence bond crystals for
an odd integer $N_0$. Finally, we briefly report results on exotic states 
for non-integer numbers of bosons per site in one-dimensional lattices. 
\\
\\
{PACS number: 05.30.Jp, 05.50.+q, 75.10.Jm, 75.45.+j}
\end{abstract}

\begin{multicols}{2}

\narrowtext

Recently, Greiner et al. reported the observation of superfluid and
Mott insulating states in optical lattices\cite{bloch01}.
The experiment convencingly demonstrates the feasibility of 
investigating many-body states of cold atoms 
by varying laser intensities of optical lattices
\cite{Fisher89,Jaksch98}.
Because of controlable properties of optical lattices, 
it is now widely accepted that many strong-coupling limits which
are not accessible in solid-state systems can be  
investigated in optical lattices.
Furthermore, since optical lattices are free from imperfections,
an accurate comparison between theories and future experiments appears to be 
likely. 
Consequently, we believe that     
studies of spin-one atoms in optical environments might also create
new opportunities for understanding some fundamental issues
in quantum magnetism.
This in turn should be  
useful to possible applications toward quantum information storage and
processing.

Spin correlated condensates have been recently investigated in experiments 
\cite{Stamper-Kurn98,Stenger98};
two-body scatterings between spin-one atoms were found to result in either 
ferromagnetic 
or antiferromagnetic condensates\cite{Ho98,Ohmi98,Law98}.
Due to spin-phase separation there also 
exist a variety of novel condensates 
which carry distinct fractionalized topological 
excitations\cite{Zhou01,Demler01,Zhou02}.
Interesting spin correlated fractional quantum hall states were  
also analyzed \cite{Ho02,Read02}.

For spin-one bosons with antiferromagnetic interactions in high 
dimensional optical lattices, 
spin correlated Mott insulating states 
were 
investigated very recently\cite{Demler01}. 
Spin singlet and nematic Mott insulators  
were pointed out for interacting spin-one bosons in strong-coupling
and intermediate coupling limit respectively.
In this Letter we examine spin correlations in 
low dimensional Mott 
insulating states.
We show that in the strong-coupling limit in one-dimensional optical 
lattices, 
there are two distinct Mott states depending on the "even-odd" parity of 
numbers of spin-one bosons per 
site ($N_0$). When $N_0$ is an even number, 
the ground state is a non-degenerate spin singlet Mott state and
the lowest lying elementary 
spin excitations correspond to
spin $S=2$ ones; for an odd integer $N_0$, the ground state is a dimerized 
valence bond crystal which has
a two-fold degeneracy and an elementary
spin excitation in this state corresponds to a spin-one state.
We also report results for one-dimensional lattices with non-integer 
numbers of atoms
per site and square 
lattices.
To get oriented, let us first consider the following coupled "mesoscopic"  
spinor condensates.

{\bf Two coupled condensates of spin-one atoms}
Consider atoms such as sodium ones
which interact with each other via 
antiferromagnetic
interactions\cite{Ho98,Ohmi98,Law98}.
Assuming all atoms condense to a single one-particle orbital state, 
we have shown that the dynamics of this condensate is 
characterized by
a constrained quantum rotor model\cite{Zhou01,Demler01,Zhou02}.
The effective Hamiltonian
of two coupled spinor condensates (at site A and B) of spin-one bosons 
with antiferromagnetic interactions 
can therefore be expressed as

\begin{eqnarray}
&& {\cal H}_{AB}=
\sum_{k=A,B} \frac{{\bf S}_k^2}{2 I} +\frac{{N}_k^2}{2C} 
-N_\alpha \mu +{\cal H}_{hop},
\nonumber \\
&&{\cal H}_{hop}= -t {\bf n}_A \cdot {\bf n}_B \cos(\chi_A -\chi_B).
\label{0D}
\end{eqnarray}

In Eq.\ref{0D}, at each site $k$ two unit vectors
${\bf n}_k$ (defined on a two-sphere) and $e^{i\chi_k}$ (defined on
a unit circle) characterize
the 
orientation of an $O(3)$- and $O(2)$-
rotor respectively; 
they are conjugate variables of the spin operator ${\bf S}_k$ and 
atom number operator $N_k$,
i.e. $[{\bf 
S}_{k\alpha},{\bf n}_{k\beta}]=$
$-i\hbar \epsilon_{\alpha\beta\gamma}{\bf n}_{k\gamma}$, $[N_k, 
\chi_k]=i\hbar$.
The total spin of a condensate at site $k$ is therefore the 
angular momentum of the $O(3)$-rotor at the same site.

${\cal H}_{hop}$ originates from hopping of bosons from one site to 
other and conserves the total spin of two coupled condensates.
$t$ is equal to $N_0 t_0$; $t_0$ is the hopping integral of an individual 
boson. $t$ can be varied by changing
laser intensities in optical lattices\cite{bloch01}. 
$E_s={(2I)}^{-1}$
$=4\pi \hbar^2 (a_2-a_0)\rho_0/3M N_0$ and
and $E_c={(2C)}^{-1}$
$=4\pi \hbar^2 (2 a_2+a_0) \rho_0/3M N_0 $
characterize the spin and "charge" gaps in the excitation spectrum.
They depend on the atomic mass $M$, the atomic number density $\rho_0$ and  
two-body scattering lengths
$a_{2,0}$ in the total spin $S=2, 0$ channels for two spin-one cold atoms. 
And for sodium atoms,
$a_2$ exceeds $a_0$ by a few percent\cite{Ho98}.
$N_0(\gg 1)$ is the number of bosons per site
and $\mu$ is the chemical potential. 

Finally,
the Hilbert space of the Hamiltonian in Eq.1 is furthermore subject to a 
constraint $(-1)^{S_k 
+n_k}=1$
so that the corresponding microscopic wave functions are symmetric under 
the interchange of two bosons; 
$S_k, n_k$ are the eigenvalues of
${\bf S}_k, {N}_k$ respectively
(see section {\bf II B} in Ref.\cite{Zhou02} for more 
discussions on the constraint).
We will consider a situation where $E_s$ is much smaller than $E_c$
because the difference between $a_2$ and $a_0$ is much smaller than $a_0$ 
as for sodium atoms; 
furthermore in the strong- 
coupling limit which interests us, $t$ is much smaller than both $E_s$ and 
$E_c$.

When
the energy gap $E_{c}(\gg E_s)$ is much larger than the hopping energy 
$t$, bosons are localized at each site;
and if there are $N_0=2k$ bosons at each site
and the hopping is absent ($t=0$),
the ground
state for each isolated site is an $S=0$ state where each pair of atoms 
forms a spin singlet, following Eq.1 and the constraint.
This limit was also discussed in \cite{Law98}.
Excitations have to be spin
$S=2,4,6...$ states and are gapped by an energy of order of $E_s$.
Therefore, two {\em weakly}
coupled condensates also have a spin singlet ground state and all 
excitations 
are gapped by 
either
$E_s$ or $E_c$. 
Particularly, there will be 
no excitations at an energy scale much lower than $E_s$.
In lattices of all dimensions, ground states in this limit 
are spin singlet 
Mott insulating states as pointed out in a previous 
work\cite{Demler01} (see Fig.1a)).

The situation for an odd number of particles per site is more involved.
When the hopping is absent, the ground state for each individual site
is a spin-one state with a three-fold degeneracy.
Eq.1 and the constraint further indicate that excitations are states with 
spin $S=3,5,...$ which
are gapped 
from the spin-one ground state.
So at energies smaller than $E_s$,
each individual spinor condensate can be treated as a spin-one {\em 
pseudo-particle}.
For two decoupled sites,
the ground state has a nine-fold
degeneracy, corresponding to states with total spin $S=0, 1, 2$. 

A detailed calculation, taking into account virtual hoppings between two 
sites, shows that the effective exchange interaction
(for $E_s$ smaller than $E_c$) is
${\cal H}_{AB}=- J \sum_{S=0,1,2} \alpha_S P_S({\bf S}_A +{\bf S}_B)$.
Here $P_S({\bf S}_A+{\bf S}_B)$ is the projection 
operator of a total spin $S$ state of the two condensates A and B.
$J=t^2/E_c$ is the strength of the exchange energy and is much 
smaller than $E_{c,s}$;

\begin{eqnarray}
&& \alpha_S=\frac{\gamma_1(S)}{1-4c_s}+\frac{\gamma_2(S)}{1+2c_s}
+\frac{\gamma_3(S)}{1+8c_s},
\label{alpha}
\end{eqnarray}
where $\gamma_1(0)=1/6$, $\gamma_3(0)=2/15$;
$\gamma_3(1)=1/10$; $\gamma_2(2)=2/15$, $\gamma_3(2)=7/150$;
$\gamma_2(0)=\gamma_1(1)=\gamma_2(1)=\gamma_1(2)=0$.
$c_s=E_s/E_c$;
and $\alpha_0$ is always larger than $\alpha_{1,2}$.

Eq.\ref{alpha} indicates that 
two coupled spinor condensates, each of which has $N_0=2k +1$ atoms again 
have a spin 
singlet ground state.
The lowest lying spin excitations
correspond to $S=2$ ones, whose
energies in this case are equal to $\epsilon(S=2)$ 
$=J\big(\alpha_0-\alpha_2\big)$.
In addition, there are 
excitations of spin $S=1$, with energies 
$\epsilon(S=1)$ $=J(\alpha_0-\alpha_1)$ and  
$\epsilon (S=1) > \epsilon (S=2)$.
In the following we will focus on ground states and excitations 
in low dimensional optical lattices with odd numbers of bosons per site
in the strong-coupling limit.
The "even-odd" feature emphasised in this letter is absent in the phase 
diagram for spinless bosons discussed 
in Ref.\cite{Fisher89}.

{\bf 1D lattices with odd numbers of bosons per site}
For the study of one-dimensional
lattices, we generalize the two-site Hamiltonian in Eq.1 to 

\begin{equation}
{\cal H}_{1D}=-J\sum_{<ij>} \alpha_S P_S({\bf S}_i + {\bf S}_{j}).
\label{ol}
\end{equation}
The sum is over $<ij>$, all pairs of nearest neighboring condensates in 
one-dimensional lattices. 
Eq.\ref{ol} is equivalent to the bilinear-biquadratic spin chain model
${\cal H}/J=\cos\eta \sum_{<ij>}{\bf S}_i \cdot {\bf S}_j $ 
$+ \sin\eta \sum_{<ij>}({\bf S}_i \cdot {\bf S}_j)^2$,  
if $\eta$ satisfies 
$\tan\eta=
(\alpha_1-\frac{1}{3}\alpha_2-\frac{2}{3}\alpha_0)/
(\alpha_1-\alpha_2)$ and the sign of $\sin\eta$ is chosen to be 
the same as that of $\alpha_1-\frac{1}{3}\alpha_2-\frac{2}{3}\alpha_0$.

The ground state of the bilinear-biquadratic 
Hamiltonian depends on $\eta$ as suggested in
\cite{Affleck87}. 
The $\eta=0$ point represents a usual $S=1$ Heisenberg antiferromagnetic
spin chain which was also studied
in\cite{Haldane83}. 
When $\eta$ varies between $-3\pi/4$ and
$-\pi/4$, one expects that the ground state of coupled condensates
should be a dimerized valence-bond crystal, or {\em DVBC}.

As $c_s$ defined after Eq.2 varies from $0$ to $+\infty$,   
$\eta$ varies from $\eta_0$ to $-\pi/2$; $\eta_0$ $=-\pi/2 -\arctan 
1/2$. For spin-one bosons such as sodium atoms, 
$c_s$ is much smaller than unity and our calculation shows that
$\eta (\approx \eta_0)$ is precisely within the dimerized regime.
This is further supported by a direct mapping between the problem of 
interacting spin-one bosons and a quantum dimer model
(see below and Ref.16 for more discussions);
it is also consistent with destructive interferences
between $Z_2$ instantons in one-dimensional lattices for an odd number of 
atoms per 
site\cite{Zhou03}.
A straightforward calculation yields that the energy per site of 
a {\em DVBC} state is 
$\frac{E}{J}=-[ \frac{5}{9} \big(\alpha_0 +\frac{\alpha_2}{2}\big)
+\frac{\alpha_1}{6}]$,
with $\alpha_{0,1,2}$ given in Eq.\ref{alpha}.

In a dimerized configuration shown in Fig.1b), 
each two neighboring condensates are in a spin singlet state,
or form a {\em valence bond};
this results in a valence bond crystal doubling the 
period of the 
underlying optical lattices.
A {\em DVBC} state can be considered 
as a projected BCS 
pairing state
of {\em pseudo-particle}s introduced above. For one-dimensional lattices 
of $2M$ ($M \rightarrow +\infty$) sites with a periodic 
boundary condition,
a DVBC state reads
(up to a normalization factor) 

\begin{eqnarray}
&& |g>={\cal P}_{G.G.}{\cal P}_{2M}
\exp\big( \sum_{\alpha=0,1} g_{\alpha}\Psi^+(Q_0=\alpha\pi)\big) 
|0>
\label{DVBC}
\end{eqnarray}
where $\Psi^+(Q_0)
=\frac{1}{V_T}\sum_Q  
\frac{h(Q)}{2\sqrt{3}}$
$[A^+_{Q,1}A^+_{-Q+Q_0,-1}+$
$A^+_{-Q+Q_0,1}A^+_{Q,-1}$ $
-A^+_{Q,0}A^+_{-Q+Q_0,0}]$.
We have introduced $A^+_{Q,m}$ as a bosonic creation operator of a
spin-one 
{\em pseudo-particle} with 
a crystal momentum $Q$ and
$m=0,\pm 1$;
$h(Q)=\exp(iQ)$ and
$Q \in [-\pi,\pi]$.     
$g_0=1, g_1=\pm 1$ 
correspond to two-fold 
degenerate   
DVBC states\cite{finite}. 
${\cal P}_{G.G.}$ is a generalized Gutzwiller projection
to forbid a double or higher occupancy of {\em pseudo-particle}s at each 
site and ${\cal P}_{2M}$ is a projection into $2M$-particle states.
Eq.\ref{DVBC} represents a projected superposition of singlet 
pairing 
states in $(Q,-Q)$ 
and $(Q,\pi-Q)$ channels.
It appears to be plausible to observe interference between these 
two channels\cite{PWA}.

\begin{figure}
\begin{center}
\epsfbox{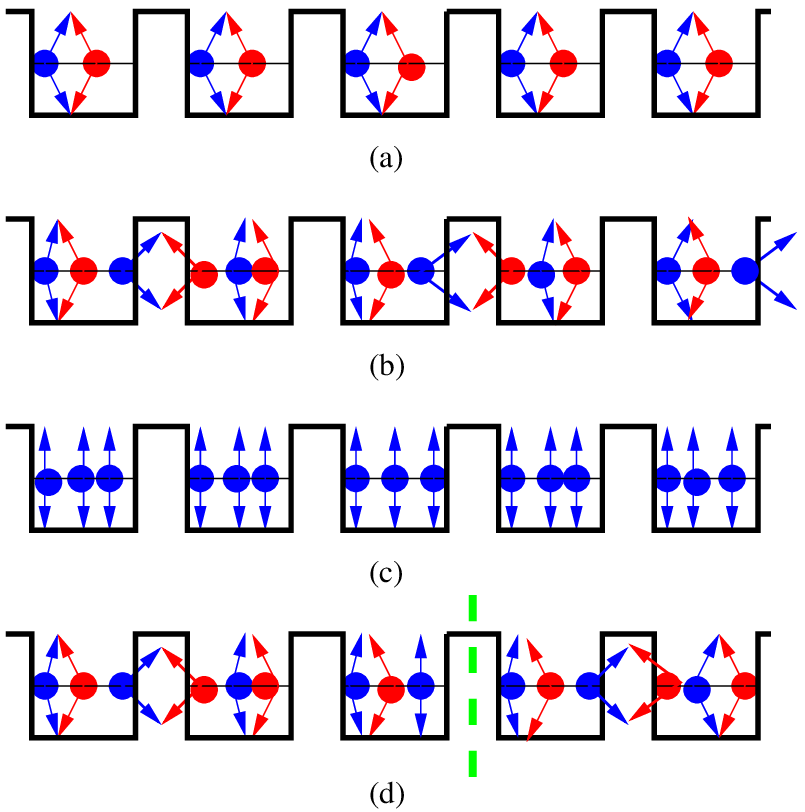}
\leavevmode
\end{center}
Fig.1 Schematic of microscopic wave functions for different 
one-dimensional Mott insulating states of 
spin-one atoms. a) A spin singlet Mott state for an even number of 
atoms per site; b) a DVBC Mott state for an odd number of atoms per site;
as a reference, we also show in c) a nematic Mott insulating state
which is stable only in high dimensional lattices\cite{Demler01};
d) a spin-one excitation (kink-like) in a {\em DVBC} state; the location
of the kink is indicated by a green dashed line.
In a), b), and d), each pair of blue and red dots represents a spin 
singlet state of two spin-one atoms. In c), d), 
a unpaired blue dot carrying an arrow 
pointing toward the direction of 
${\bf \Omega}$ represents an atom in a spin-one 
state specified by ${\bf \Omega}=(\theta,\phi)$:
$\frac{1}{\sqrt{2}}\sin\theta \exp(-i\phi)|1,1>
-\frac{1}{\sqrt{2}}\sin\theta \exp(i\phi)|1,-1>+\cos\theta|1,0>$;
in c), localized atoms at each site "condense" in an identical spin-one 
state.
\end{figure}

The {\em DVBC} state breaks the crystal translational symmetry and has a 
twofold 
degeneracy; one of the degenerate states is characterized by the following 
correlation functions
($i\neq j$)

\begin{eqnarray}
&& <{\bf S}_i \cdot {\bf S}_j >=
-\frac{1}{4}\big( 1 + (-1)^j \big)
\delta_{j, i+1}, 
\nonumber \\ 
&& < {\bf S}_i \cdot {\bf S}_{i+1}
{\bf S}_j \cdot {\bf S}_{j+1} >=\frac{1}{16} 
\big(1- (-1)^i\big) \big(1- (-1)^j \big).
\label{correlation}
\end{eqnarray}
Eq.\ref{correlation} indicates 
long-range order in valence bonds;
the alternating feature 
between site $i=2k+1$ and $i=2k$ 
in the correlation 
function 
reflects a two-fold degeneracy of  
DVBC states\cite{Nematic}.

An $S=1, S_z=m$ excitation can be created by the following product
operator 
\begin{equation}
C^+_{\gamma, m}={\cal P}_{G.G.} A^+_{\gamma,m}\prod_{\xi \in 
{\cal C}_{\gamma}}
[\Psi_{\xi}^+ +\Psi_{\xi}]
\label{kink}
\end{equation}
where $\Psi^+_\xi=\frac{1}{\sqrt{3}}(A^+_{i,1}A^+_{j,-1}+
A^+_{i,-1}A^+_{j,1}-A^+_{i,0}A^+_{j,0})$ is a 
creation
operator of a valence bond at link
$\xi$ which connects two neighboring sites $i$ and $j$;
$A^+_{\gamma,m}$ is the creation operator of a pseudo-particle or a 
collective state at site 
$\gamma$ with spin
$S=1$, $S_z=m$. 
Finally, the product is carried over all links $\xi$ along 
path ${\cal C}_\gamma$ which starts at site $\gamma$ and ends at the 
infinity;
furthermore, ${\cal C}_\gamma$ is chosen such that the first link 
along the path is 
occupied by a valence bond.
Eq.\ref{kink} also represents
a spin-one domain wall soliton(of $S_z=0,\pm 1$) located at site 
$\gamma$ 
in the DVBC as shown in Fig.1d).  
Taking into account the hopping matrix of the domain wall 
${\cal T}_{ij}=-\frac{1}{{3}}\alpha_0[\delta_{i,j-2}+\delta_{i,j+2}]$
along one-dimensional lattices, we obtain the following 
band structure for spin-one excitations:

\begin{equation}
{E(Q_x)}=J\alpha_0\big(1 -\frac{2}{3}\cos 2 {Q_x} \big)
\label{band}
\end{equation}
with $Q_x$ defined as a crystal momentum of the excitation,
$-\pi/2 < Q_x < \pi/2$.
These excitations are purely magnetic and involve no extra atoms,
i.e. they are "neutral".


{\bf 1D lattices with non-integer numbers of bosons per site}
When the number of bosons per site deviates from an integer,
extra bosons can hop along one-dimensional lattices and 
form a superfluid state. 
The situation with $N_0=2k\pm \epsilon$, $\epsilon \ll 1$ is particularly
simple. In this case, $2k$ spin-one bosons localized at each site are
in a spin singlet state;
extra bosons added to or removed from the system tend to condense in a 
state with zero crystal momentum.
However, hyperfine spin
dependent scatterings destroy spin long-range order of condensed
spin-one bosons in one dimension as pointed out in \cite{Zhou01,Zhou02}.
The ground 
state therefore is spin disordered but phase coherent, similar to that in 
one-dimensional traps.
Interactions between bosons furthermore lead to usual Bogolubov 
quasi-particles
with a sound-like spectrum as well as gapped spin-one spin-wave 
excitations.

To facilitate discussions on the case of $N_0=2k+1\pm \epsilon$,
we introduce, besides spin-one "neutral" excitations,
"charged" spinless domain-wall excitations. A "charged" excitation with a 
positive or negative 
charge ($\pm $) is 
defined by creation operator $B^+_{\gamma, \pm}$ and

\begin{equation}
C^+_{\gamma, m}= \psi_{\gamma,-m} B^+_{\gamma,+}  
=\psi^+_{\gamma,m} 
B^+_{\gamma,-}
\label{frac}
\end{equation} 
where $\psi^+_{\gamma,m}(\psi_{\gamma,m})$ is a creation (annihilation) 
operator 
of a spin-one
atom with $S_z=m$. 
Following Eqs.\ref{DVBC},\ref{frac},
as it is added to one-dimensional
DVBCs, a spin-one boson fractionalizes into
a spin-zero domain wall (with an extra atom thus "charged") and a spin- 
one "neutral" domain wall because of a two-fold degeneracy 
at $N_0=2k+1$. 
This is reminiscent of the physics in conducting ploymers\cite{Heeger88}.
At $N_0=2k+1\pm \epsilon$,
both types of domain walls are free quasi-particles and
charged domain walls can condense at the bottom
of a band similar to that in Eq.\ref{band}, as a consequence of
spin-"charge" separation in one-dimensional optical lattices.
Spin-charge separation was previously pointed out 
for correlated electrons in antiferromagnets\cite{Anderson87}.

{\bf Disordered states in square lattices}
In the limit when $E_{c,s}$ are much larger than $t$,
our analysis suggests that the
constrained quantum rotor model 
(in Eq.\ref{0D})
generalized to square lattices 
should be equivalent to 
an Ising gauge theory
in (2+1)D in a quantum spin-disordered limit (for detailed 
discussions on the mapping, see \cite{Zhou03}).
The effective Hamiltonian is
(in an arbitrary unit) 
${\cal H}_{p.c.}=-\Gamma \sum \sigma_x - 
\sum_{\Box}\prod_{\Box}\sigma_z$,
where $\sigma_x$ and $\sigma_z$ are Pauli matrices defined at each link 
and
$\Box$ stands for an elementary plaquette in 
square lattices in this case. The Hilbert space of
${\cal H}_{p.c.}$ is subject to a local constraint $\prod_+ 
\sigma_x =(-1)^{N_0}$ at each site, 
with the product carried over all links connected with that
single 
site. In (2+1)D, this Ising gauge field model
${\cal H}_{p.c.}$ is dual to an Ising model(frustrated when $N_0$ is 
an odd integer) in a 
transverse field 
and also represents a parity conserved quantum dimer 
model\cite{Wegner71,Kogut79,Sachdev91,Senthil01,Moessner01,Sachdev00}.
The frustrated Ising gauge field theory
has been employed 
to investigate electron bond ordering  
\cite{Sachdev91,Senthil01,Moessner01,Sachdev00,Read91}.

As $t$ becomes much smaller than $E_s$, $\Gamma$ approaches infinity. 
By examining the model straightforwardly, one finds that the 
ground state of ${\cal 
H}_{p.c.}$ for $N_0=2k $ is a non-degenerate spin singlet; 
this is consistent with previous discussions on Mott states for an even 
number of atoms.
For $N_0=2k +1$ and $\Gamma \gg 1$,
the ground state breaks no continuous symmetries but does break the lattice
translational symmetry.
It has a four-fold degeneracy and 
supports no gapless modes. 
This result implies that in a square lattice in a quantum spin disordered 
limit, the ground 
state for 
an odd number of bosons per site   
should be a valence-bond crystal (columnar-type) state.
Two dimensional valence-bond crystals have also been proposed in a few 
recent models 
on strongly- 
correlated electrons \cite{Senthil01,Moessner01,Sachdev00} and in 
earlier works\cite{Haldane88,Read89,Read91,Sachdev91,Kivelson88}.

The ground state $|g>$ on a torus of $2M\times 
2M$ sites should be topologically identical with one of the 
following configurations

\begin{equation}
{\cal P}_{G.G.}{\cal P}_{4M^2} \exp\big(\sum_{\eta,\xi=0,1}
g_{\eta\xi}\Psi^+({\bf Q}_0=(\eta\pi,\xi\pi)) \big)|0>.
\label{2dvbc}
\end{equation}
In Eq.\ref{2dvbc}, $g_{11}=0$.
$g_{00}=\pm g_{10}=1$, $g_{01}=0$, $h({\bf Q})=\exp(i{\bf Q}\cdot {\bf 
e}_x 
)$ and $g_{00}=\pm g_{01}=1$, $g_{10}=0$, $h({\bf Q})=\exp(i{\bf Q}\cdot 
{\bf e}_y)$ 
correspond to four-fold degenerate DVBC states in square lattices
($\Psi^+({Q}_0)$ is defined after Eq.4).

An examination of Eq.\ref{2dvbc} indicates that both 
"charged" spinless solitons
and spin-one "neutral" solitons 
similar to those introduced in one-dimensional lattices interact with each 
other via a
linear confining potential 
in a two-dimensional DVBC.
The ground state therefore supports excitations of $S=2,1$ bound states
of soliton anti-soliton pairs 
unlike in one-dimensional DVBC states.

We should emphasis here the important differences 
between the physics of Mott insulating states of interacting spin-one 
bosons  
and of Heisenberg antiferromagnets (unfrustrated).
First, for interacting spin-one bosons, the Hilbert space at each site 
involves 
different spin states($S=1,3,5,...$ for an odd $N_0$ much larger than 
unity) unlike in Heisenberg 
antiferromagnets; this leads to a unique Ising gauge symmetry
in the problem of interacting spin-one bosons\cite{Zhou01,Zhou02}, nematic 
Mott states 
in high dimensional 
optical lattices\cite{Demler01} and dimerization in low dimensional Mott 
insulating 
states.
Second, phase transitions between nematic  
and spin singlet Mott insulators for spin-one bosons can be
conveniently investigated by changing exchange interactions 
between two sites or practically by tuning
laser intensities in optical lattices. 
This opportunity appears to be absent in simple unfrustrated 
Heisenberg anti-ferromagnets.

Finally, we notice that two-fold degenerate states in one
dimensional optical lattices 
could be qubits of a quantum computer.
I would like to thank F. D.M 
Haldane for many stimulating discussions on the BLBQ model,
E. Demler for an early collaboration on a 
related 
subject and the 
year 2001 workshop on "beyond BEC"
at Harvard-MIT center for ultracold atoms for its hospitality.
I am also grateful to Bianca Y. L. Luo for encouragement and 
support.
This work is supported by the foundation FOM, in the Netherlands
under contract 00CCSPP10, 02SIC25 and by NWO Projectruimte 00PR1929.

\end{multicols}


\begin{thebibliography}{99}
\bibitem{finite}
Due to the propagation of virtual kink-anti kink pairs along 
lattices,
a finite geometry such as a ring lifts the degeneracy; the splitting is 
exponentially small as the perimeter of the ring increases.
\bibitem{Nematic}
The DVBC has no gapless modes.
\end{thebibliography}
\end{document}